\documentclass[journal,twoside,web]{ieeecolor}
\usepackage{generic}
\usepackage{cite}
\usepackage{amsmath,amssymb,amsfonts}
\usepackage{algorithmic}
\usepackage{graphicx}
\usepackage{algorithm,algorithmic}
\usepackage{hyperref}
\hypersetup{hidelinks=true}
\usepackage{makecell}
\usepackage{textcomp}
\def\BibTeX{{\rm B\kern-.05em{\sc i\kern-.025em b}\kern-.08em
    T\kern-.1667em\lower.7ex\hbox{E}\kern-.125emX}}
\begin{document}
\title{Charting the Path Forward: CT Image Quality Assessment – An In-Depth Review}
\author{Siyi Xun, Qiaoyu Li, Xiaohong Liu, Guangtao Zhai, Mingxiang Wu, Tao Tan
\thanks{This work is supported by Science and Technology Development Fund of Macao (0021/2022/AGJ), Macao Polytechnic University Grant (RP/FCA-15/2022), and Macao Polytechnic University Grant (RP/FCSD-01/2022). (\textit{Siyi Xun and Qiaoyu Li contributed equally to this work.}) (\textit{Corresponding authors: Tao Tan, Mingxiang Wu, and Xiaohong Liu})}
\thanks{Siyi Xun and Tao Tan are with the Faculty of Applied Sciences, Macao Polytechnic University, Macao, 999078, China (e-mails: p2214963@mpu.edu.mo; taotan@mpu.edu.mo).}
\thanks{Qiaoyu Li is with the Faculty of Languages and Translation, Macao Polytechnic University, Macao, 999078, China (e-mail: p2108988@mpu.edu.mo).}
\thanks{Xiaohong Liu and Guangtao Zhai are with Shanghai Jiao Tong University, Shanghai, 200240, China (e-mail: xiaohongliu@sjtu.edu.cn; zhaiguangtao@sjtu.edu.cn).}
\thanks{Mingxiang Wu is with the Department of Radiology, Shenzhen People's Hospital, Luohu, Shenzhen, 518020, China (e-mail: szmxwu@outlook.com).}
}

\maketitle

\begin{abstract}
Computed Tomography (CT) is a frequently utilized imaging technology that is employed in the clinical diagnosis of many disorders. However, clinical diagnosis, data storage, and management are posed huge challenges by a huge volume of non-homogeneous CT data in terms of imaging quality. As a result, the quality assessment of CT images is a crucial problem that demands consideration. The history, advancements in research, and current developments in CT image quality assessment (IQA) are examined in this paper. In this review, we collected and researched more than 500 CT-IQA publications published before August 2023. And we provide the visualization analysis of keywords and co-citations in the knowledge graph of these papers. Prospects and obstacles for the continued development of CT-IQA are also covered. At present, significant research branches in the CT-IQA domain include Phantom study, Artificial intelligence deep-learning reconstruction algorithm, Dose reduction opportunity, and Virtual monoenergetic reconstruction. Artificial intelligence (AI)-based CT-IQA also becomes a trend. It increases the accuracy of the CT scanning apparatus, amplifies the impact of the CT system reconstruction algorithm, and creates an effective algorithm for post-processing CT images. AI-based medical IQA offers excellent application opportunities in clinical work. AI can provide uniform quality assessment criteria and more comprehensive guidance amongst various healthcare facilities, and encourage them to identify one another's images. It will help lower the number of unnecessary tests and associated costs, and enhance the quality of medical imaging and assessment efficiency. 

\end{abstract}


\section{Introduction}
\label{sec:introduction}
\IEEEPARstart{C}{omputed} Tomography (CT) is one of the imaging modalities that is most frequently utilized in clinical practice ~\cite{1}. CT images are crucial for the clinical diagnosis of disorders of the neurological system, cardiovascular system, organs in the chest and abdomen, bones, and joints because of imaging speed and high contrast between bones and soft tissues. As a result, CT imaging is a crucial tool for routine physical examinations and illness screening. 

Consequently, a significant volume of CT images is produced every day. Massive non-homogeneous CT images, however, makes clinical diagnosis as well as the management and storage of medical data more challenging~\cite{2}. It is easy for low-quality images to impede clinical diagnosis, leading to incorrect diagnoses. It is also very difficult to determine images of minimal acceptable quality that are sufficient for clinical usage or in a research setting. Simultaneously, the different quality has a significant impact on the acceptance of CT images across various medical facilities.It will lead to additional radiation exposure for patients, raise the risk of recurrent cancer, and waste societal resources. Therefore, CT image quality assessment (IQA) is of great significance.

CT-IQA refers to the process of determining the extent of image distortion by examining and analyzing the fundamental characteristics of CT images (contrast, clarity, etc.) and other factors affecting the quality of CT images (patient position, artifact, etc.). It is a crucial area of research for computer vision, medical image processing, and related disciplines. Soares~{et al.}~\cite{31} conducted the first study on CT-IQA in 1994. During the initial phase of CT-IQA development, radiologists' subjective assessments served as the primary basis for the assessment procedure. With the development of computer technology and digital image processing, the objective assessment method using algorithms and mathematical models comes into being. In recent years, the method of automated CT-IQA using AI models has gradually entered the field of vision of researchers, and has become an important research direction.

\begin{figure*}[!t]
\centerline{\includegraphics[width=1\linewidth]{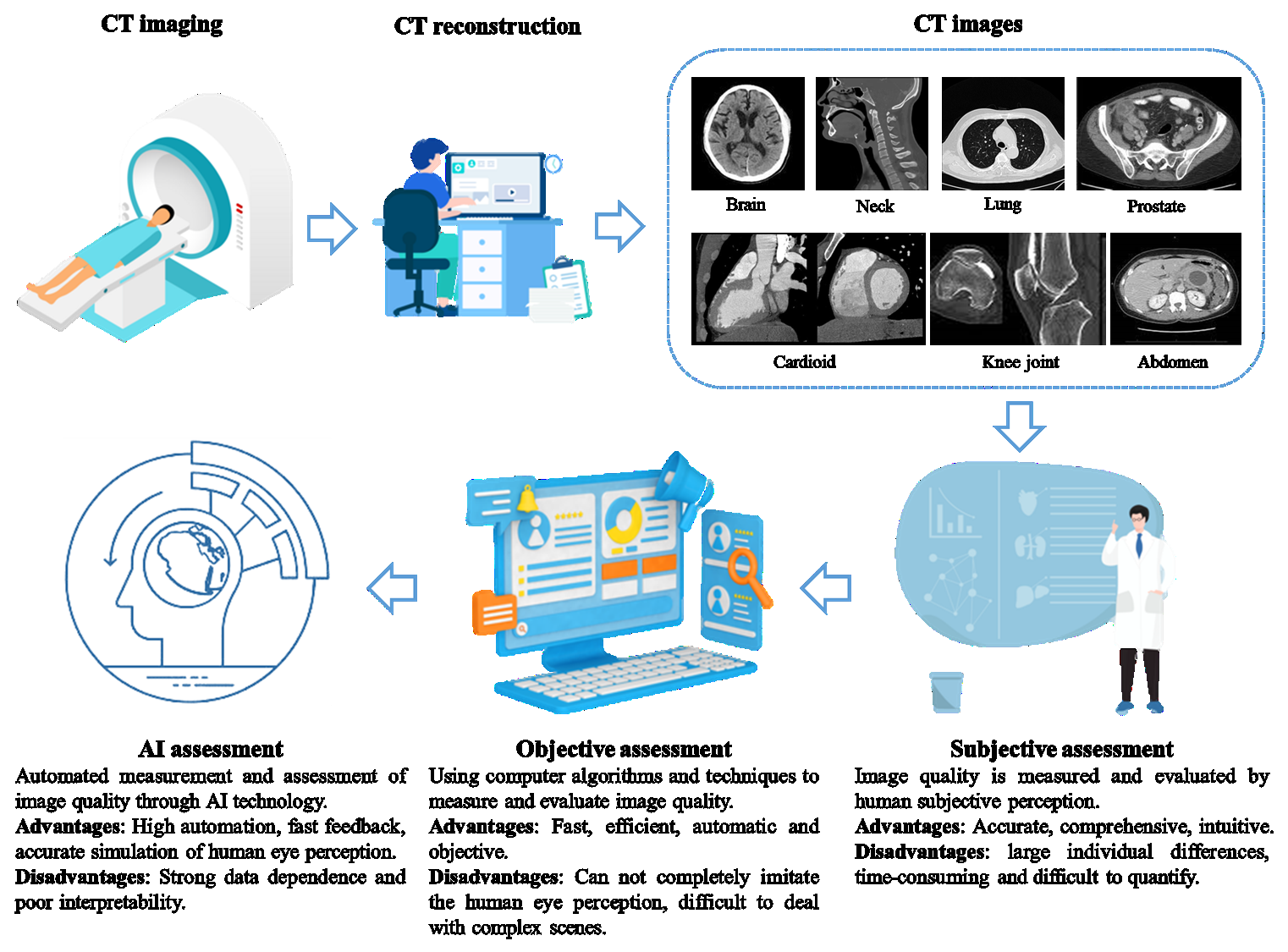}}
\caption{Schematic diagram of CT imaging, reconstruction, and quality assessment. Different types of CT images are obtained by CT imaging and reconstruction, and CT-IQA methods of subjective assessment, objective assessment, and automatic assessment based on AI are extended.}
\label{Fig1}
\end{figure*}

This paper reviews the history, current research topics, and development of CT-IQA. It also discusses the prospects and obstacles for CT-IQA's further development, as illustrated in Figure~\ref{Fig1}. By utilizing the visualization graph analysis of literature linked to CT-IQA, we are able to describe the historical timeline of CT-IQA by using the turning point in the field's development as the anchor point. In the meantime, we investigate the new issues in the field by utilizing the co-occurrence and co-citation relationships between literature. Lastly, we talk about the difficulties and potential paths ahead for the CT-IQA domain.

This is how the rest of the article is organized. First, the development of CT-IQA and the standard metrics to assess CT-IQA models are covered in Section 2. In Section 3, we used \textit{CiteSpace} to analyze the papers we had retrieved and summarized them based on key branches, future trends, and the development process. Section 4 discusses the aforementioned review. Section 5 provides our conclusions.

\section{Background}
\subsection{CT Imaging's Development}

CT imaging technology has been widely employed in many different fields since its introduction in the 1970s. Researchers have enhanced CT scanners from various perspectives in the past few decades~\cite{3}. Table~\ref{tab1} displays the technical characteristics and development process of the five-generation CT scanners. From the early low resolution, lengthy scanning times to the contemporary high-speed, high resolution, and low radiation dose equipment, CT imaging is constantly evolving.

Different CT imaging machines have varying resolution, noise, contrast, radiation dose, and scanning speeds due to the quantity and arrangement of detectors, the focus, and the energy of X-ray sources. As a result, the quality of the CT images generated by the scan differs. The quality of CT images has significantly increased due to ongoing technological advancements, making them essential tools in contemporary medicine and playing a significant role in clinical diagnosis.

\begin{table}
\caption{Five-generation CT scanners' technical characteristics and development process.}
\label{table}
\setlength{\tabcolsep}{3pt}
\begin{tabular}{|p{60pt}|p{170pt}|}
\hline
\textbf{Development} & \textbf{Technical characteristics} \\ \hline
1970-1980s: First generation CT scanners & The human body is scanned by rotation using X-ray sources and detectors. Afterwards, several two-dimensional slices are combined into three-dimensional images using computer methods that have a low resolution and a lengthy scan duration. \\ \hline
1980-1990s: Second generation CT scanner & Improved resolution and rotation speed. Spiral CT technology was also introduced to expedite the scanning of the complete body. \\ \hline
1990-2000s: Third generation CT scanner & The technology of multi-row detectors was introduced. The scanning time can be significantly decreased with this technology's ability to gather numerous slices at once, and the scanner's spatial resolution is also enhanced. \\ \hline
2000-2010s: Fourth generation CT scanner & The resolution and scanning speed were further enhanced by adding more detector lines. Furthermore, more sophisticated image reconstruction algorithms have been developed, enabling a more precise reconstruction of the human body's interior structure. \\ \hline
Since 2010s: Fifth generation CT scanner & More sophisticated radiation dose control technology as well as higher time resolution were implemented. Furthermore, a growing number of transportable CT scanners have been developed to do scans outside of clinical settings. \\ \hline

\hline

\end{tabular}
\label{tab1}
\end{table}

\subsection{Image Quality Assessment}

IQA is one of the key technologies in the digital image processing domain. By analyzing and exploring the features of images, it can evaluate the degree of image distortion (that is, the quality of images)~\cite{4}.  IQA can be categorized as Full Reference IQA (FR-IQA), Reduced Reference IQA (RR-IQA), and No Reference IQA (NR-IQA) depending on whether a reference image is used as the assessment standard~\cite{5}. Depending on the subject of IQA judgment process, it can be divided into subjective assessment and objective assessment.

Subjective assessment is based on visual perception and subjective consciousness. Double stimulus damage grading, double stimulus continuous quality grading, and single stimulus continuous quality grading are the three most widely used techniques~\cite{6}. While the subjective assessment accurately captures the image quality, handling a huge number of images is a challenge. Therefore, numerous objective assessment methods for evaluating image quality have been developed. Table~\ref{tab2} displays typical metrics of objective assessment.

As Artificial Intelligence (AI) has grown, many techniques and metrics for IQA based on AI have been suggested and enhanced by researchers in recent years. Le~{et al.}~\cite{14} achieved accurate NR-IQA by using a 5-layer Convolutional Neural Networks (CNN). In order to get better performance, Bosse~{et al.}~\cite{15} employed end-to-end networks to share the same network with full reference and no reference assessment during training. To determine the final assessment result, Bianco~{et al.}~\cite{16} separated the detected images into sub-regions and summed the expected scores of each region. The NR-IQA methods based on deep learning promotes the development of IQA, which even outperform the results of FR-IQA with the ongoing advancements in deep learning technology.

\begin{table}
\caption{Common objective methods for assessment and metric meanings.}
\label{table}
\setlength{\tabcolsep}{3pt}
\begin{tabular}{|p{50pt}|p{175pt}|}
\hline
\textbf{Metric} & \textbf{Meaning} \\ \hline
HaarPSI~\cite{7} & The Haar wavelet decomposition coefficient is used to determine the local similarity between the ideal image and the image that needs to be assessed.  \\ \hline
LPIPS~\cite{8} & The depth feature is used as a perceptual metric to assess how similar two images are perceptually. \\ \hline
MS-SSIM~\cite{9} & Combining the outcomes of several structural similarity computations at various resolutions (scales) yields the final assessment value. \\ \hline
MSE~\cite{10} & The divergence between observed and true values is measured by the mean square error between true and projected values. \\ \hline
SSIM~\cite{11} & The degree of image distortion is determined by an in-depth assessment of the sample image's brightness, contrast, and structure. \\ \hline
PSNR~\cite{11} & It represents the relationship between the signal's maximal potential value (power) and the distortion noise power that compromises the signal's quality of representation. \\ \hline
VIF~\cite{12} & Based on the idea of natural scene statistics and image signal extraction by human visual system, information entropy and mutual information are utilized to assess image quality. \\ \hline
VSI~\cite{13} & The gradient modulus and chrominance features are utilized to build additional complementary features, and the image saliency feature map is used to calculate the image distortion. \\ \hline

\hline

\end{tabular}
\label{tab2}
\end{table}

\subsection{IQA of Medical Images}

Medical image is an important branch of digital image domain. Currently, the assessment of medical images still depends primarily on radiologists' subjective opinions. But in addition to being tedious and time-consuming, subjective quality assessment is also susceptible to the evaluator's subjective biases. Therefore, it is very important to develop efficient automated medical IQA methods.

Numerous studies has extended natural image IQA techniques to medical image IQA~\cite{17, 18, 19}. These techniques do, however, nevertheless have some limitations because of the unique nature of medical imaging. Lei~{et al.} conducted medical image IQA of various modes using deep learning and machine learning techniques~\cite{14, 20, 21}. Mason~{et al.}~\cite{23} examined 10 FR-IQA methods, while Stepien~{et al.}~\cite{24} analyzed NR-IQA methods for Magnetic Resonance Imaging (MRI) in order to investigate the effect of IQA methods on MRI images. Furthermore, Kastryulin~{et al.}~\cite{25} assessed the use of 35 distinct IQA techniques in MRI using the assessment criteria of noise, contrast, and artifact presence. These techniques included full reference methods and no reference methods.

There isn't a flawless image that can be referred to as the ``gold standard'' in medical image IQA. Therefore, many researchers believe that the most effective IQA method for medical images is NR-IQA~\cite{26}. Nonetheless, a key difficulty in the present medical image IQA is how to assess only based on the information and characteristics of the medical image itself.

\subsection{CT-IQA Metrics}

For the CT-IQA model, the Video Quality Experts Group (VQEG) proposed four metrics that can confirm the degree of correspondence between objective assessment results and subjective assessment outcomes in order to quantify the consistency between model test results and subjective assessment: Pearson Linear Correlation Coefficient (PLCC), Spearman Rank-order Correlation Coefficient (SROCC), Kendall Rank-order Correlation Coefficient (KROCC), and Root Mean Square Error (RMSE).

\subsubsection{PLCC}

PLCC assesses the accuracy of the IQA model by calculating the Mean Opinion Score (MOS) and the linear correlation with the objective score after following  regression. The value falls between -1 and 1. The two sets of data have no relationship at all when the PLCC value is zero; on the other hand, a PLCC value of 1 or -1 denotes a complete positive or negative correlation. Equation~\ref{eq:sample1} calculates the logistic function of the objective score for nonlinear regression,

\begin{equation}\label{eq:sample1}
	p(Q) = \beta _{1} [\frac{1}{2} - \frac{1}{1+e^{(\beta _{2} (Q-\beta _{3} ))} }]+\beta _{4}Q+\beta _{5} 
\end{equation}

where \textit{Q} represents the original objective mass score, and $\beta _{1}$, $\beta _{2}$, $\beta _{3}$, $\beta _{4}$, $\beta _{5}$  are the fitting parameters and \textit{p} is the objective score after regression operation. Equation~\ref{eq:sample2} calculates the PLCC,

\begin{equation}\label{eq:sample2}
	PLCC = \frac{\sum_{i = 1}^{N}(s_{i}-\bar{s})
(p_{i}-\bar{p} ) }{\sqrt{\sum_{i=1}^{N} (s_{i}-\bar{s})^{2} \sum_{i=1}^{N} (p_{i}-\bar{p})^{2} } } 
\end{equation}

where $s_{i}$ and $p_{i}$ represent MOS and objective score of images \textit{i} respectively, and $\bar{s} $and $\bar{p} $ represent average MOS and objective score respectively.

\subsubsection{SROCC}

SROCC uses a linear correlation analysis of the rank sizes of two target arrays to quantify the monotonicity of IQA prediction. The value falls between 0 and 1. There is consistency between the two sets of data if the performance value is 1. Equation~\ref{eq:sample3} calculates the SROCC,

\begin{equation}\label{eq:sample3}
	SROCC = 1-\frac{6\sum_{i = 1}^{N} d_{i}^{2} }{N(N^{2}-1 )} 
\end{equation}

where \textit{N} represents the number of samples, and $d_{i}^{2}$ represents the difference between the MOS ranking and the objective score ranking of image \textit{i}.

\subsubsection{KROCC}

KROCC, like SROCC, is used to measure the monotonicity of IQA model predictions. The correlation is higher the larger the value. Equation~\ref{eq:sample4} calculates the KROCC,

\begin{equation}\label{eq:sample4}
	KROCC = \frac{2(N_{c} - N_{d}) }{N(N-1)} 
\end{equation}

where \textit{N} represents the number of samples, $N_{c}$ is the number of ``harmonious pairs'' (refers to the two sample observations with the same order of variable size, that is, the order of the \textit{x} level is the same as the order of the \textit{y} level, otherwise it is referred ``disharmonious'') in the data set, and $N_{d}$ is the number of non-harmonious pairs in the data set.

\subsubsection{RMSE}

The RMSE evaluates the consistency of the IQA model's predictions by calculating the absolute error between the objective score and the MOS score. Better model performance is shown by the value being closer to 0. Equation~\ref{eq:sample5} calculates the RMSE,

\begin{equation}\label{eq:sample5}
	RMSE = \sqrt{\frac{1}{n}\sum_{i=1}^{N} (s_{i} - p_{i})^{2}  } 
\end{equation}

where $s_{i}$ and $p_{i}$ represent the MOS value and objective score of image \textit{i} respectively.

These metrics assess the CT-IQA model's performance from various perspectives. It is important to remember that in real-world applications, combining various metrics to analyze in tandem is essential for obtaining a more comprehensive assessment of the model.

\section{In-depth Domain Analysis}

\begin{figure}[!t]
\centerline{\includegraphics[width=1\linewidth]{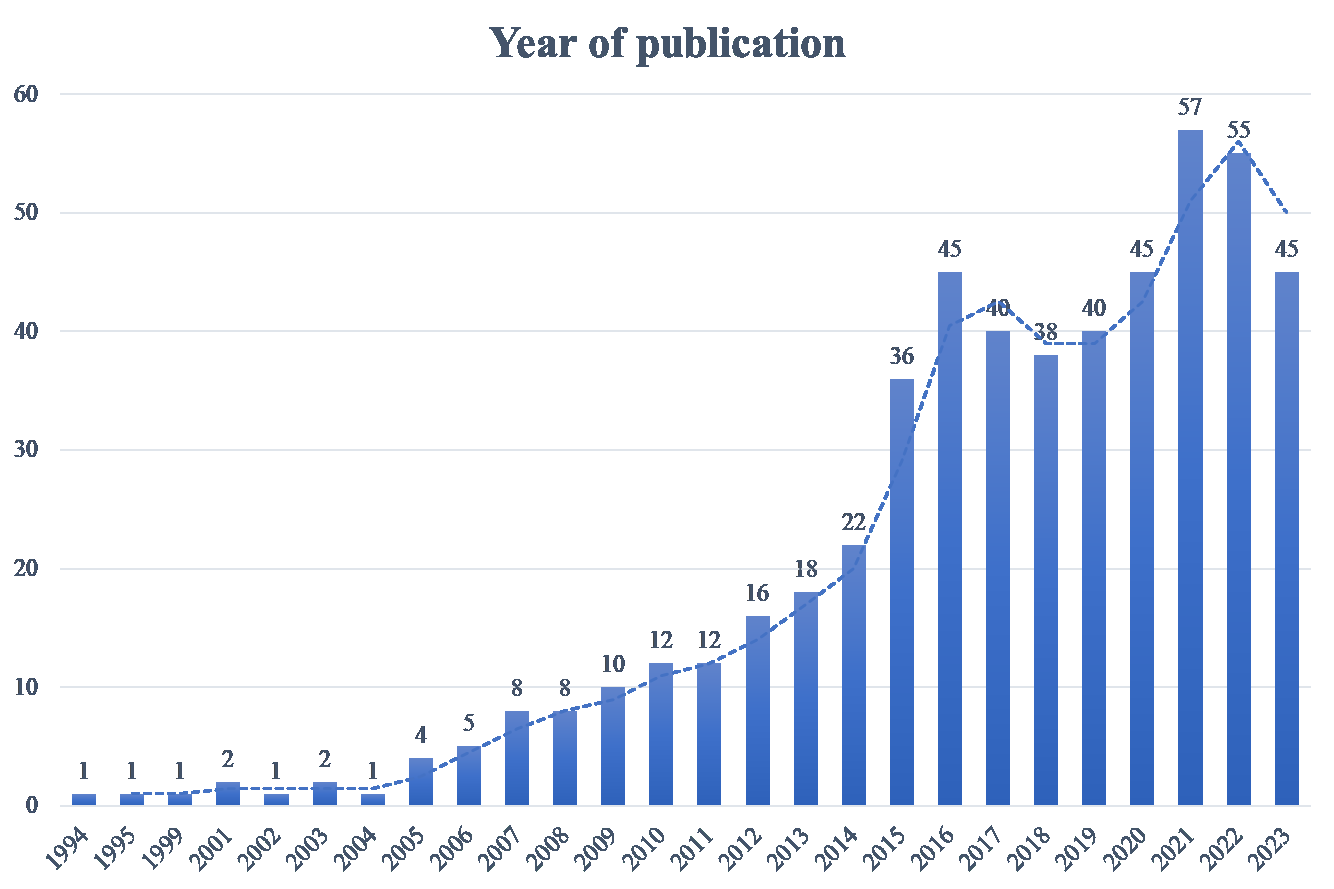}}
\caption{Year distribution of relevant papers.}
\label{Fig2}
\end{figure}

The primary source of information for our visual analysis of the ``CT image quality assessment'' topic is the \textit{Scopus} database search results. We focus on essential terms such as ``CT'', ``medical image'', and ``Image Quality Assessment (IQA)'' in the retrieval process. Simultaneously, to improve the quality and applicability of the visual analysis graph, we removed unrepresentative paper categories—like conference proceedings—and disregarded publications that had no connection to the aforementioned keywords. This methodology resulted in the creation of an extensive dataset that contained 525 research articles from 1994 to August 2023. The chronological distribution of these papers is shown in Figure~\ref{Fig2}. 

Given the limitation of \textit{CiteSpace}~\cite{27, 28, 29, 30} in integrating results from diverse databases, our review process incorporated significant findings from additional repositories, namely \textit{Google Scholar}, \textit{IEEE Xplore}, \textit{PubMed}, \textit{Semantic Scholar}, and \textit{Springer}, to ensure a comprehensive overview of the pertinent fields. Furthermore, we conducted a meticulous comparison between the references in the selected papers from the aforementioned databases and the co-occurrence atlas produced by \textit{CiteSpace}. This was undertaken to manually validate the scientific accuracy and comprehensiveness of the atlas.

\textit{CiteSpace} version 6.2.R6 Advanced was used in this investigation. The major goals of this study are to describe the origins and development of CT-IQA, investigate the pivotal trends within its development, elucidate the knowledge structure in this domain, and ascertain the presence of any paradigm shifts. To achieve these objectives, we utilized keywords and citation bibliographies as nodal points to construct two distinct graphs: a time-zone graph of co-occurring keywords and a cluster graph of co-citation. To exhibit more simply and intuitively, we merged keywords with the same meaning, eliminated the display of irrelevant words, and merged the same literature with different citation formats during the production process.

Initially, we rendered the co-occurrence keyword graph map of the dataset's published information, integrating temporal segments from 1994 to 2023 (refer to Figure~\ref{Fig3}). This graphical representation facilitates an analysis of the evolving research focal points in the CT-IQA domains from a longitudinal perspective. Subsequently, a burst detection of keywords was conducted, employing a sensitivity threshold of 1.0 and a minimum burst duration of one year. This process identified 24 keywords exhibiting significant surges (designated as `burst keywords'), depicted as red dots in Figure~\ref{Fig3}. These burst words are instrumental in discerning the research emphases of various time periods. For an in-depth examination and discussion of Figures~\ref{Fig3} and Figure~\ref{Fig4}, Section 3.1 should be consulted.

Secondly, we visualized the references in the publication information of the dataset and clustered the nodes in the initially obtained co-citation graph (see Figure~\ref{Fig5}). Figure~\ref{Fig5} illustrates four subsets of nodes with the densest node connections in the network, representing the four core clusters. We employed the ``log-likelihood ratio'' algorithm to assign cluster labels. The emergence of core clustering enables us to highlight pivotal literature in various research directions from the perspective of subdomain (see section 3.2 for detailed analysis). In addition, we also detected the burst of the co- citation. With a sensitivity of 1.0 and a minimum duration of 2 years, 11 references with the strongest reference bursts were found (see Figure~\ref{Fig6}), which are denote red dots in Figure~\ref{Fig5}. The literature with high burst citations can also indicate the research hotspots of each period.

Finally, we analyze the structure variation of the cluster of co- citation. The analysis considered a citation span of 5 years, with a focus on the more recent co-citations. This is demonstrated by the line segments marked with purple and red arrows in Figure~\ref{Fig5}. This provides a reference for us to analyze whether the knowledge structure of the CT-IQA domain has undergone a paradigm shift and predict future development trends (see section 3.3 for detailed analysis).

\subsection{CT-IQA Development}

\begin{figure*}[!t]
\centerline{\includegraphics[width=1\linewidth]{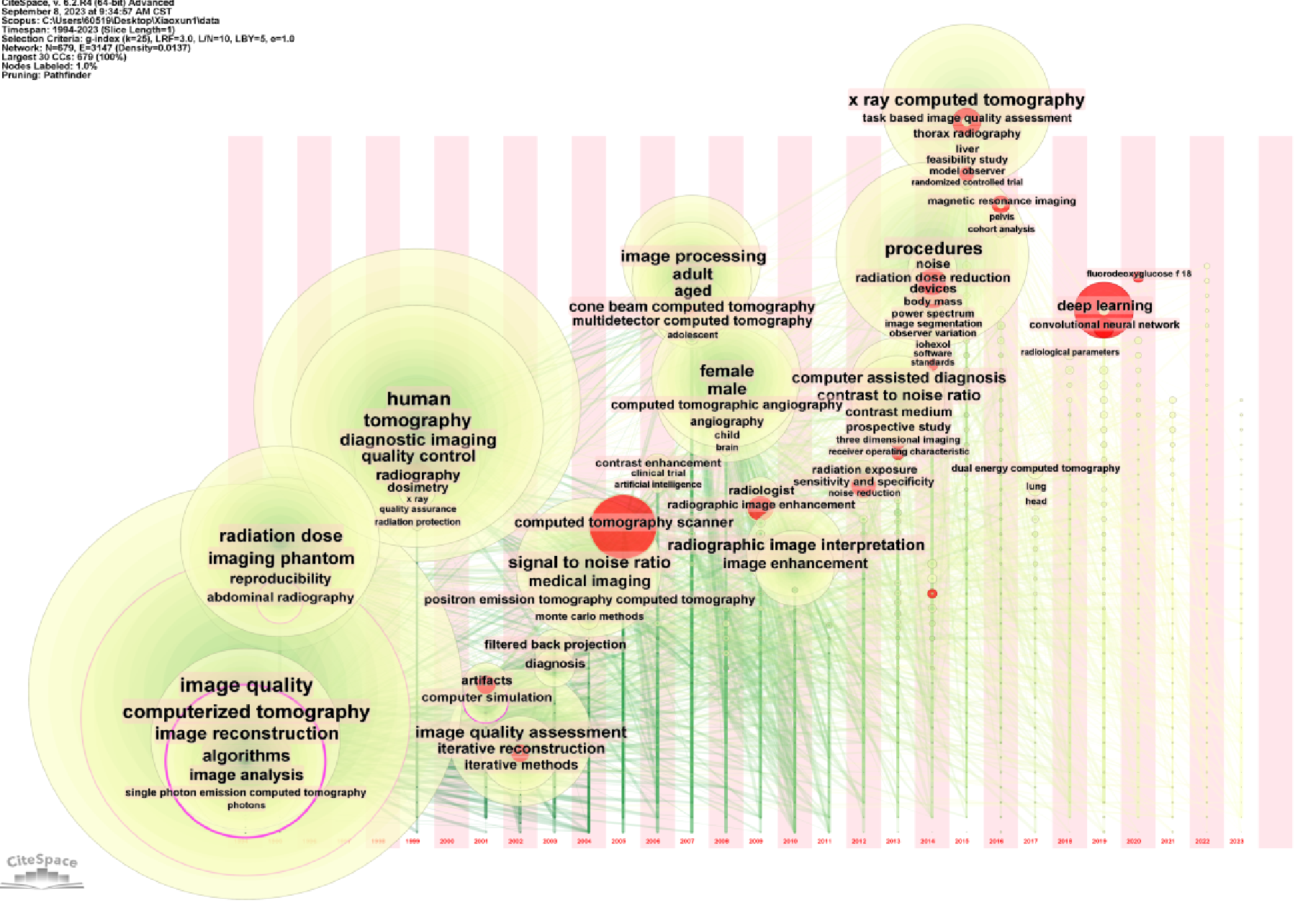}}
\caption{A time-zone diagram representing the evolution of CT-IQA research from 1994 to 2023. The horizontal axis of the graph map indicates the years, segmented into annual intervals, depicted as alternating pink and white color blocks. Nodes in this diagram represent keywords, symbolized by circles, with the size of each circle corresponding to the keyword's frequency of occurrence. Keywords are centrally displayed within each circle, labeled in black font on a pink foundation. Additionally, a purple ring encircles keywords demonstrating high centrality, whereas a red dot signifies keywords with notable bursts, as identified in Figure 4. The presence of connections between nodes is derived from the co-occurrence relationship among keywords. Specifically, when two keywords simultaneously appear within the publication information of an article, a connecting line is formed. The gradation of the line's color (from darker to lighter shades) illustrates the evolution of the co-occurrence relationship over time (from the earlier part to the later part).}
\label{Fig3}
\end{figure*}

\begin{figure*}[!t]
\centerline{\includegraphics[width=1\linewidth]{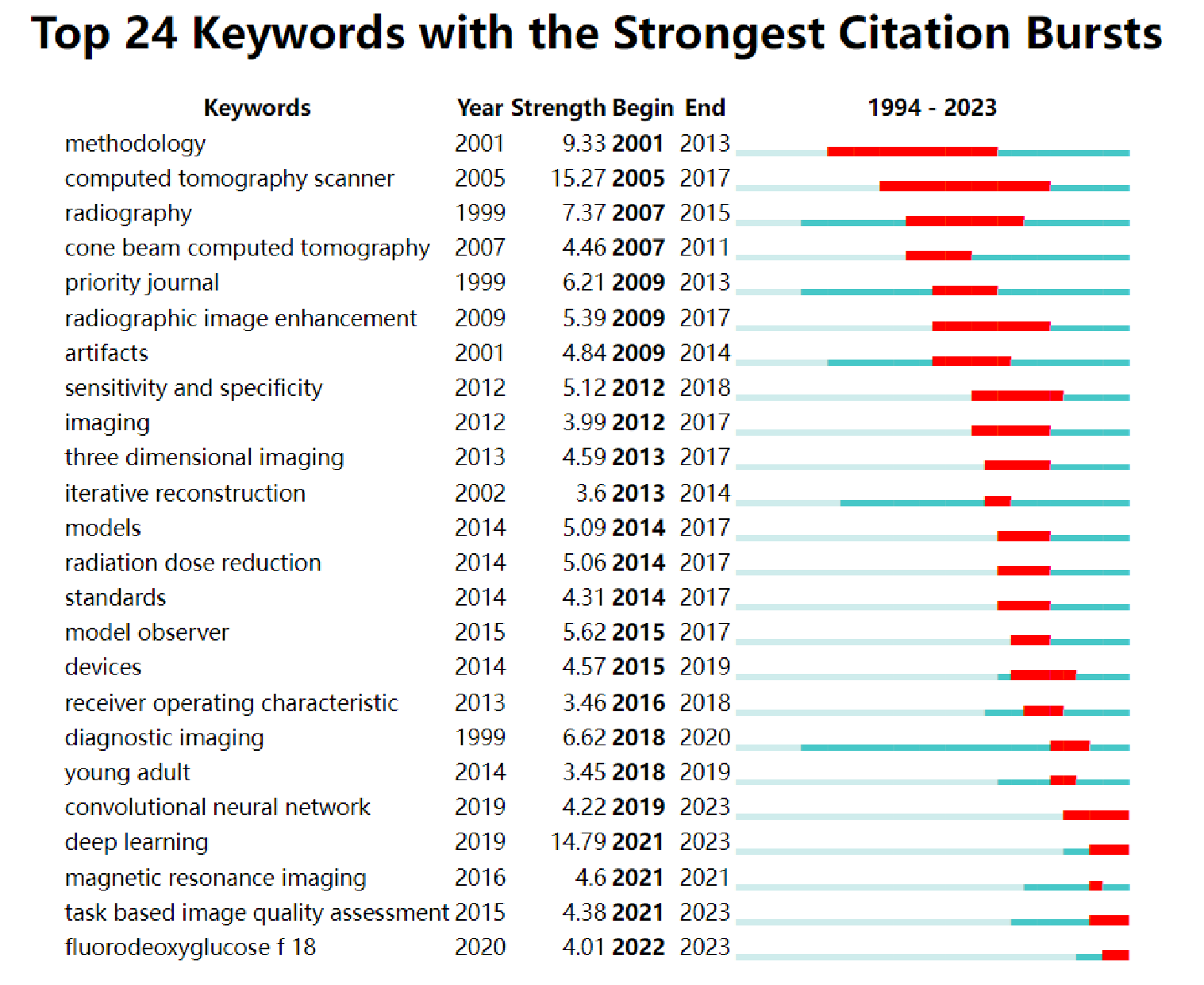}}
\caption{Keywords with the strongest citation bursts (from left to right: burst term detected, year of first appearance of the term, burst intensity, year of burst commencement, year of burst conclusion. The sixth column provides a visual representation of the term's burst from 1994 to 2023, with the entire blue line depicting the period from 1994 to 2023. The starting point of the dark blue line represents the year of the term's initial appearance, while the starting point of the red line indicates the year of the burst's initiation. The light blue time period has not yet materialized).}
\label{Fig4}
\end{figure*}

The evolution of CT-IQA during the last 30 years is analyzed and summarized in this part based on the key development nodes shown in the timeline diagram (Figure~\ref{Fig3}) and burst word diagram (Figure~\ref{Fig4}). 

\subsubsection{The Beginning of CT-IQA}

The CT imaging was first presented in 1972 and was progressively implemented. People's awareness of CT image quality is growing as a result of the ongoing advancements in CT imaging technology and equipment. For the first time, Soares~{et al.}~\cite{31} assessed the quality of CT images in a research study published in 1994. They also examined the effects of various attenuation correction techniques on the noise characteristics of reconstructed single photon emission computed tomography (SPECT) images. The phantom is then used for the first time by Suleiman~{et al.}~\cite{32} to investigate how radiation exposure affects the quality of CT and X-ray images at various US locations. Radiation safety and quality control in Asia and East Asia are examined by Oresegun~{et al.}~\cite{33}. Since then, scientists have begun to assess, investigate, and improve the quality of CT images from a variety of angles and domains, including imaging technology and image processing methods.

\begin{figure*}[!t]
\centerline{\includegraphics[width=1\linewidth]{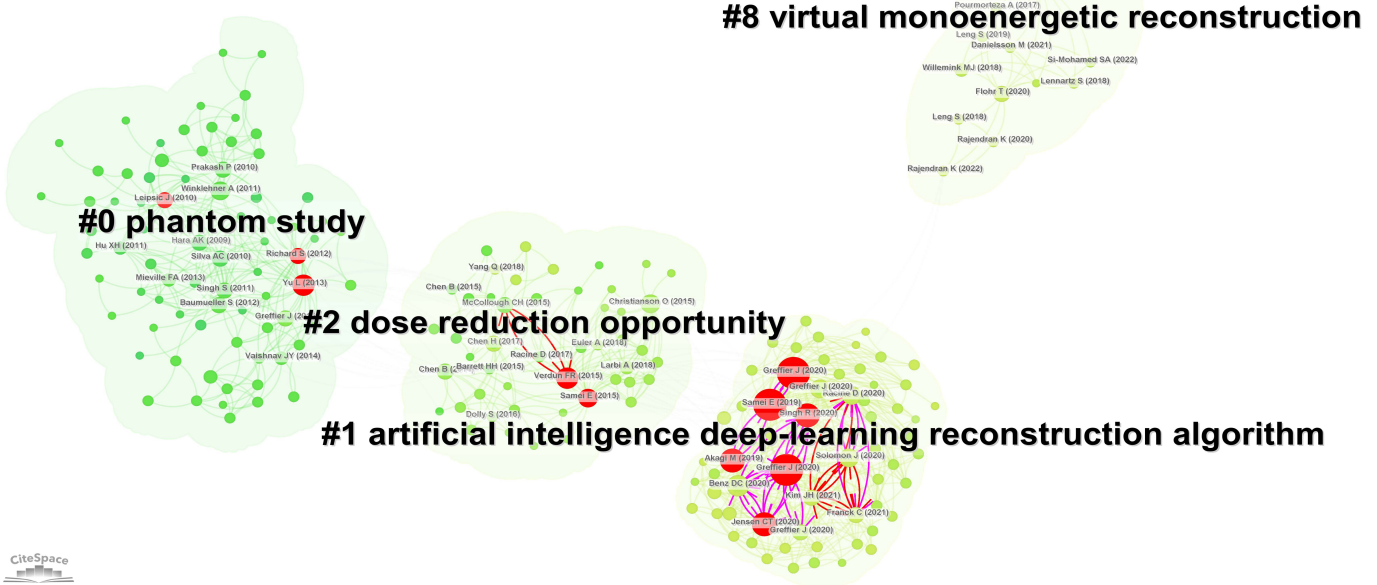}}
\caption{Cluster graph depicted the development of CT-IQA (The nodes in this graph represent references, denoted by circular shapes, and grey text denotes the author and year information of frequently co-citation. The four core clusters in the figure are categorized in ascending order of scale as follows: \#0 phantom study, \#1 artificial intelligence deep-learning reconstruction algorithm, \#2 dose reduction opportunity, \#8 virtual monoenergetic reconstruction. Purple lines signify pre-existing connections among various nodes, while red lines indicate newly established connections).}
\label{Fig5}
\end{figure*}

\begin{figure*}[!t]
\centerline{\includegraphics[width=1\linewidth]{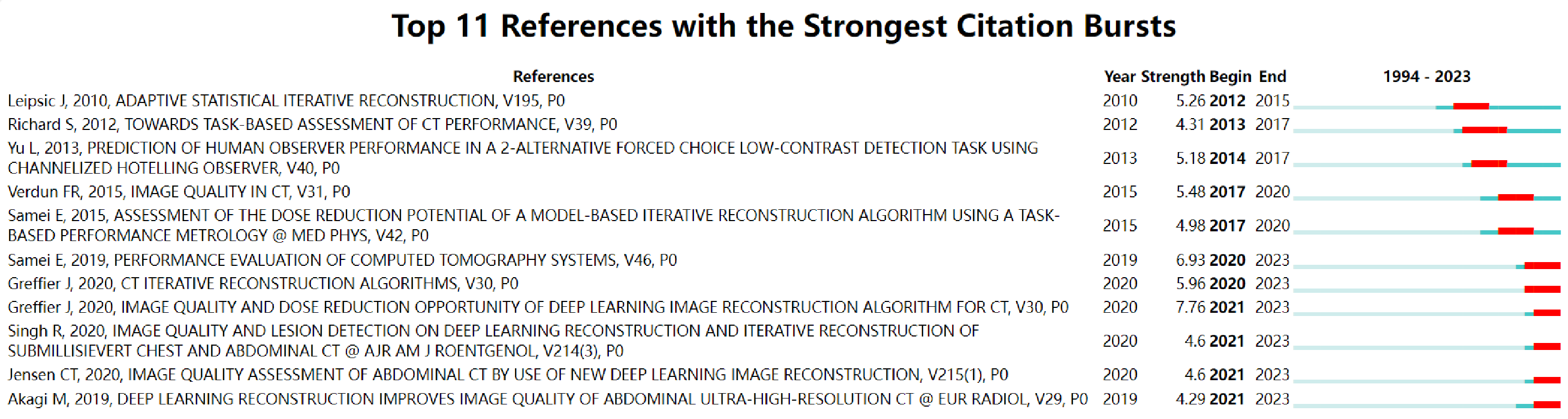}}
\caption{References with the strongest citation bursts (same explainability as Figure~\ref{Fig4}, except that this figure shows nodes as references).}
\label{Fig6}
\end{figure*}

\subsubsection{Artifacts in CT-IQA}

As time went on, first burst word ``artifacts'' began to emerge in 2001 and remained popular for five years, from 2009 to 2014. Artifacts are abnormal changes in density that appear on an image that do not correspond to the actual anatomical structure. The presence of artifacts may cause a physician to make a mistaken diagnosis, which would reduce diagnosis accuracy. Regarding the sources of artifacts, Damini~{et al.}'s~\cite{34} investigation of artifacts in Coronary Computed Tomography Angiography (CCTA) using Dual-Source CT (DSCT) revealed that calcified plaques and cardiac phase dislocation were the primary influencing variables of these artifacts. Dmitry~{et al.}~\cite{35} offer a technique for quantifying the degree of artifacts in CT images, which can aid in fine-tuning parameters and algorithms to produce high-quality images via quantitative analysis. In addition, researchers will find fresh study paths and ideas from Edward~{et al.}'s thorough examination and explanation of the causes and reduction approaches of artifacts in CT images~\cite{36}.

\subsubsection{IQA in CT Imaging Systems}

Then came ``iterative reconstruction'', and then the most intense and persistent keywords, ``computed tomography scanner'', which started in 2005 and kept growing until 2017. The two burst words mentioned above are crucial parts of the CT imaging system. To create 2D or 3D images with spatial resolution and contrast, a set of projection data from a CT scanner is first processed through reconstruction filtering, reconstruction calculations, and other steps in the CT imaging system. 

As a result, the researchers investigated how various CT scanners affected the caliber of CT images. Z-axis flying focus technology has been shown by Flohr~{et al.}~\cite{37} to efficiently suppress artifacts and improve CT images when it comes to image reconstruction. Cone Beam CT (CBCT) scan image technical and quality assessment is carried out by Xu~{et al.}~\cite{38}, who also guided the updating of technical methods to guarantee image quality. Recently, Patzer~{et al.}~\cite{39} conducted quantitative and qualitative IQA on related images using a novel type of CT scanner. The investigation revealed that this imaging mode has the potential to become an important imaging modality for clinical assessment.

Meanwhile, for the CT image reconstruction algorithm, Vardhanabhuti~{et al.}~\cite{40} have conducted a comparison of the image quality of popular reconstruction techniques for CT images: Filtered Back Projection (FBP), Adaptive Statistical Iterative Reconstruction (ASIR), and Model-Based Iterative Reconstruction (MBIR). The outcomes demonstrate the superior noise reduction and enhanced image quality of MBIR. The most recent research on task-based CT Iterative Reconstruction (IR) IQA techniques was compiled by Vaishnav~{et al.}~\cite{41}, who also assessed the degree of dose reduction attained by pertinent IR algorithms.

The researchers also conducted a comparison analysis for several CT systems. Over six years, Maria~{et al.} ~\cite{42} investigated quality control techniques for various systems by examining the characteristics and variability of six distinct CT scanner systems from four different manufacturers. Additionally, a CT system IQA approach has been developed by several researchers. For instance, Allert~{et al.}~\cite{43} conducted a thorough evaluation of the CT system from many perspectives, including contrast and spatial resolution, and based on this, they constructed an automated assessment program. Additionally, Wilson~{et al.}~\cite{44} provide an IQA technique for CT systems. It is capable of accurately measuring the system's tube current modulation and IR performance. Using imQuest software, Greffier~{et al.}~\cite{45} assess the performance of four CT systems, as well as the effects of FBP, two generations of IR algorithms, and two detection tasks for high mass and tiny calcification in the liver on image quality. These investigations have the potential to significantly enhance both the quality of CT scans and the system's performance.

\subsubsection{IQA of CT Image Processing Algorithm}

People are no longer content with imaging machine research with the advent of CT imaging. ``radiographic image enhancement'' first surfaced in 2009 and kept growing rapidly until 2017. In an effort to enhance image quality through post-processing technology, researchers have progressively started to investigate the image post-processing methodology. To enhance the contrast and brightness of CT images, Zohair~{et al.}~\cite{46} improved the ratio restriction adaptive histogram equalization and suggested a straightforward contrast enhancement technique. An iodine contrast enhancement tool was studied by Peter~{et al.}~\cite{47} for use in post-processing CT images. This tool considerably improved the quality of CT images while lowering the amount of iodine needed for venous enhancement. A contrast enhancement technique for non-contrast CT is proposed by Sim~{et al.}~\cite{48}; this technology can aid in the diagnosis of brain illnesses and improve the detection of cerebral infarction areas. The quality and contrast of CT images are improved using a number of image enhancement methods.

Another crucial post-processing method for CT images is denoising. Zhu~{et al.}~\cite{49} suggest a parallel noise reduction technique that increases operating speed and enhances image quality through parameter tuning. For nondestructive assessment of industrial CT imaging systems, Lee~{et al.}~\cite{50} offer a Total Variational (TV) denoising technique, which successfully improved image noise characteristics. The quality performance of reconstruction and denoising technologies in children's abdomen CT images is investigated by Watanabe~{et al.}~\cite{51}. Post-processing techniques like filtering help to improve the quality and clarity of CT images by removing noise. Furthermore, image fusion, artifact removal, and other popular post-processing methods for CT images are described~\cite{52}. These methods enhance the visual perception and diagnostic potential of the image in addition to its quality.

\subsubsection{CT-IQA originated AI}

In the past five years, ``convolutional neural network'' and ``deep learning'' have appeared in the domain of view of relevant researchers. Among them, Deep learning (DL) intensity reached 14.79. With the development of AI, researchers are progressively integrating deep learning and neural network-based techniques into CT-IQA. Naeemi~{et al.}~\cite{53} use CNN for the first time to explore the impact of big data on CT-IQA, and then researchers began to explore the application of deep learning in CT-IQA. Using deep learning techniques, Hayashi~{et al.}~\cite{54} examined CT images with various noise levels. The findings indicated that the AI-based IQA approach will be crucial for clinical diagnosis in the future. While Lee~{et al.}~\cite{56} propose an NR-IQA method, Depth Detector IQA (D2IQA), which can automatically conduct quantitative assessment of CT image quality and even show comparable performance with FR-IQA index, Li~{et al.}~\cite{55} conduct a preliminary investigation on NR-IQA of CT images based on deep learning. A few AI-based automated CT image assessment models are displayed in Table~\ref{tab3}. The CT-IQA method, which is based on deep learning, can better capture the complex features and potential relationships of images by using deep neural networks to automatically learn the feature representation of images. This allows for the creation of high-performance prediction models that can adapt to a wide range of application scenarios, which greatly accelerates the development of CT-IQA.

\begin{table*}
\caption{Summary and comparison of some AI-based automated CT image assessment models.}
\label{table}
\setlength{\tabcolsep}{3pt}
\begin{tabular}{|p{40pt}|p{30pt}|p{50pt}|p{35pt}|p{45pt}|p{45pt}|p{80pt}|p{140pt}|}
\hline
\textbf{References} & \textbf{Modality} & \textbf{ROI} & \textbf{Algorithm} & \textbf{Loss} & \textbf{Quality level} & \textbf{Data} & \textbf{Performance} 
\\ \hline
Schwyzer~{et al.}~\cite{001} & PET/CT & Whole body & ResNet-34 & CrossEntropy & 1-4 & \makecell[l]{Private dataset: 400 \\ Train: 320 \\ Val: 40 \\ Test: 40} & \makecell[l]{BSREM beta 450, BSREM beta 600 \\ AUC = 0.9780, 0.9670 \\ Sen = 0.8900, 0.9400 \\ Spe = 0.9400, 0.9400}
 \\ \hline
Lee~{et al.}~\cite{56} & CT & Abdomen & \makecell[l]{Cascade \\ R-CNN, \\ ResNet-50} & \makecell[l]{CrossEntropy, \\ L1} & 0-10 & \makecell[l]{2016 Low Dose CT \\ Grand Challenge \\ dataset: 9228 \\ Train: 6390; Val: 1598 \\ Test: 1240} & \makecell[l]{PLCC = 0.9132 \\ SROCC = 0.9058}
 \\ \hline
Xia~{et al.}~\cite{003} & CT & Thorax & CNN & CrossEntropy & 1-2 & \makecell[l]{COVID-CT-MD \\ Train: 374 \\ Val: 175} & \makecell[l]{Ori, AD1, AD2, TV1, TV2 \\ AUC = 0.9350, 0.9150, 0.8950, 0.9450, \\ 0.9500 \\ Sen = 0.9500, 1.0000, 0.9100, 1.0000, \\ 0.9500 \\ Spe = 0.9200, 0.8300, 0.8800, 0.8900, \\ 0.9500}
\\ \hline
Wang~{et al.}~\cite{004} & CT & Thorax & \makecell[l]{Genetic \\ algorithm} & N/A & 1-5 & \makecell[l]{LIDC-IDRI: 266 \\ Train: 190; Test: 76} & SROCC = 0.9209 
\\ \hline
Imran~{et al.}~\cite{005} & CT & \makecell[l]{Thorax, \\ Abdomen} & CNN & \makecell[l]{Self- \\ supervised, \\ L2, \\ Heatmap} & 1-6 & \makecell[l]{Mayo CT data \\ Abdomen: Train 9198, \\ Test 3315 \\ Thorax: 3185} & \makecell[l]{Compare with GSSIM (Abdomen, Thorax) \\ PLCC = 0.9830, 0.9170 \\ SROCC = 0.9710, 0.9060 \\ KROCC = 0.8570, 0.7430}
\\ \hline
Gao~{et al.}~\cite{006} & CT & \makecell[l]{Thorax, \\ Abdomen} & CNN & L1 & 1-5 & \makecell[l]{Mayo CT data: 700 \\ Train: 420 \\ Val: 140 \\ Test: 140} & \makecell[l]{BF, NLM, BM3D \\ PLCC = 0.8770, 0.8731, 0.8653 \\ SROCC = 0.8296, 0.8302, 0.8184 \\ RMSE = 10.3906, 10.5805, 10.8607}
\\ \hline
Duan~{et al.}~\cite{007} & CT & \makecell[l]{Head, Thorax \\ Abdomen, Hip} & SVM & N/A & 1-2 & Private dataset: 690 & \makecell[l]{Head, Thorax, Abdomen, Hip \\ SROCC = 0.7769, 0.8661, 0.8040, 0.9094} 
\\ \hline
Gao~{et al.}~\cite{008} & CT & Thorax & CNN & N/A & 1-3 & \makecell[l]{Mayo CT data: 900 \\ Train: 720; Test: 180} & \makecell[l]{PLCC = 0.9602 \\ SROCC = 0.9548}
\\ \hline 
Li~{et al.}~\cite{55} & CT & Thorax & AlexNet & N/A & 1-5 & \makecell[l]{Mayo CT data: 1500 \\ Train: 1350; Test: 150} & \makecell[l]{PLCC = 0.9953 \\ SROCC = 0.9952}
\\ \hline 
Naeemi~{et al.}~\cite{53} & CT & \makecell[l]{Thorax, \\ Abdomen} & CNN & N/A & \makecell[l]{Phantom: 1-6 \\ Patient: 1-5} & \makecell[l]{Private dataset: 690 \\ Phantom: 81 \\ Patient: 100} & Acc = 0.9380
\\ \hline 
Hayashi~{et al.}~\cite{54} & CT & N/A & AlexNet & N/A & 1-3 & \makecell[l]{Private dataset: 90 \\ Train: 63; Test: 27} & Acc = 0.9260
\\ \hline 
Chen~{et al.}~\cite{0012} & CT & Abdomen & \makecell[l]{VLM, \\ ChatGPT} & \makecell[l]{CrossEntropy, \\MSE} & 1-5 & \makecell[l]{2016 AAPM Grand \\ Challenge Dataset: 1000 \\ Train: 800; Test: 200} & \makecell[l]{PLCC = 0.7540 \\ SROCC = 0.7480}
\\ \hline
Qi~{et al.}~\cite{0013} & PET/CT & Whole body & CNN & N/A & 1-5 & \makecell[l]{Private dataset: 173 \\ Train: 98; Val: 33 \\ Test: 32} & \makecell[l]{R1, R2 \\ Kappa = 0.7900, 0.7800}
\\ \hline

\multicolumn{8}{p{500pt}}{Acc: Accuracy; AD: Anisotropic Diffusion; AUC: Area under the ROC Curve; BF: bilateral filtering; BM3D: Block-matching and 3D filtering; BSREM: Block Sequential Regularized Expectation Maximization; N/A: Not Applicable; NLM: Non-local mean filtering; Ori: Original; R: Reviewer; ROI: Regions of interest; Sen: Sensitivity; Spe: Specificity; TV: Total Variation; Val: Validation.}

\end{tabular}
\label{tab3}
\end{table*}

\subsubsection{IQA for CT-relevant Modalities}

Furthermore, the break word ``fluorodeoxyglucose f 18'' was observed. In Positron Emission Tomography-Computerized Tomography (PET/CT), it is frequently utilized as a tracer. For PET/CT, Yin~{et al.}~\cite{57} examine the effectiveness of 13 widely used IQA methods and suggest a subjective assessment database for PET/CT images. An AI-based quality system of assessment for PET/CT images was created by Qi~{et al.}~\cite{0013}; it can analyze CT image quality quickly and generate comprehensive assessment reports. 

While conducting our investigation, we came across further multi-modal CT image studies. The IQA of the SPECT imaging system and the adjustment of imaging system parameters for detection tasks are examined by Gross~{et al.}~\cite{59}. Greffier~{et al.}~\cite{60} assessed the image quality scanned by four Dual-Energy CT (DECT) platforms as well as their spectral performance. Dillenseger~{et al.}~\cite{61} assessed CBCT systems with various points of view using both qualitative and quantitative methods. diverse types of CT imaging offer diverse image information for various application scenarios, assisting medical professionals in accurately diagnosing patients, planning treatments, and keeping track of their conditions.

In conclusion, CT-IQA plays a critical role as a foundation for clinical diagnosis and treatment. On the one hand, the accuracy of the imaging technology and parameter adjustments determines the quality of CT images. The accuracy of CT scanners is continually increasing due to the technology's ongoing development, and different CT systems' IQA procedures have been developed. Simultaneously, a range of CT imaging techniques with distinct features have been created, including CBCT, DECT, PET/CT, and other modes. Additionally, CT-IQA for various modes has progressively advanced, significantly assisting in clinical diagnosis. To achieve reasonably good quality CT images, physicians must choose the right scanning settings (such as scanning layer thickness, interval, and scanning speed) and a safe radiation dose based on clinical standards and real-world circumstances. 

On the other hand, various post-processing algorithms will also have an impact on CT image quality. Various algorithms can do 3D reconstruction and optimization, as well as suppressing CT image artifacts and enhance image contrast. The CT-IQA can be used to assess the algorithm's impact, fine-tune its parameters, choose the optimal algorithm from a range of options, and provide feedback for advancement. Simultaneously, a proper assessment approach is also required for the application effect of CT image processing technology. For instance, trustworthy CT-IQA methods are required to validate the application effects of CT image reconstruction, augmentation, denoising, and other processing algorithms.

\subsection{CT-IQA Key Topics}

The cited paper chosen for this review synthesizes the relevant studies on CT-IQA, completely expanding the scope and depth of research on this subject. We performed a knowledge structure analysis of citations over a five-year period using \textit{Citespace} in order to investigate the future development trend of this area. From this analysis, four significant research topics were identified. Themes include: ``phantom study'', ``artificial intelligence deep-learning reconstruction algorithm'', ``dose reduction opportunity'', ``virtual monoenergetic reconstruction''. In order to compile a summary of the key study areas in the CT-IQA domain, highly cited publications in each field are examined as exemplary examples, and representative literature with high burst intensity is also examined.

\subsubsection{Phantom Study in CT-IQA}

Phantom study is research projects that mimic the CT imaging process of actual human bodies using standardized simulated body models. They can assess the effectiveness of CT devices and carry out performance comparison and verification. In order to investigate the imaging quality and resolution characteristics of CT imaging systems under various dosage levels, reconstruction methods, and contrast, the top three referenced typical publications~\cite{62, 63, 64} in the cluster all used phantoms.

Through analyzing the other articles in cluster \#0, we find that along with the advance of science and technology, phantom of research has been improved. Reliability and fidelity of the phantom studies have steadily evolved from the first basic simulation of natural tissues and organs to the more intricate and realistic simulation. Based on this, a wide range of illnesses and lesions (such cancers, vascular stenosis, etc.) have been identified, which is useful in assessing how well imaging equipment can identify and diagnose lesions~\cite{65}. Phantom study has gradually advanced toward multimodal imaging as a result of the development of multimodal medical imaging technology, which allowing for a more accurate assessment of the benefits and drawbacks of each modality as well as their complementarity~\cite{66}. 

Phantom study offers carefully regulated experimental settings for CT imaging technology, algorithms, or parameters under various imaging quality performance testing and validation conditions. This will offer a scientific foundation for clinical application of promotion and use, aid in quality control and improvement, and enhance the equipment's image quality and dependability. In addition, there is a very little chance of radiation exposure throughout the study, which has emerged as a crucial area for further investigation.

\subsubsection{CT-IQA of Artificial Intelligence Deep-learning Reconstruction Algorithm}

With the continuous development of AI, researchers are using deep learning algorithms and neural networks to automatically extract and learn features by analyzing a large number of data samples. This allows them to produce a reconstruction image of the original data that is higher quality and more accurate, which enhances the quality of CT images.

Greffier~{et al.}~\cite{67} evaluate the potential of hybrid IR algorithm and deep learning image reconstruction (DLIR) algorithm for image quality and dose optimization, while Singh~{et al.}~\cite{68} compare the image quality and performance of the two algorithms on chest,  abdomen, and pelvic CT and the detection of clinically important lesions. Akagi~{et al.}~\cite{69} add MBIR for comparison. A number of studies have shown that compared with traditional reconstruction methods, DLIR can reduce noise, improve spatial resolution detectability without changing the noise texture, and has good lesion detection capabilities, while showing extraordinary potential in noise, contrast and other indicators. In addition, Jensen~{et al.}~\cite{70} quantitatively and qualitatively evaluate the DLIR algorithm in abdominal tumor enhanced CT. Similarly, DLIR effectively improves CT image quality. However, as DLIR intensity grows gradually, the reconstructed image's degree of blur increases.

Numerous research have demonstrated that reconstruction algorithms based on AI and deep learning may enhance image quality and decrease image noise in CT imaging, assisting medical professionals in making more accurate diagnosis decisions. Simultaneously, when paired with Figure~\ref{Fig6}, we discover that the papers that have maintained high bursts up to this point are all associated with the deep learning reconstruction algorithm. This further underscores the immense potential for deep learning advancement in CT-IQA.

\subsubsection{Dose Reduction in CT-IQA}

The amount of X-ray radiation that a patient receives during CT imaging is referred to  radiation dose. The contrast, resolution, noise level, and other aspects of CT image quality will all be impacted by the radiation dosage. High radiation dosages may harm human cells and tissues and raise the risk of cancer, whereas too low a dose can result in poorer-quality CT scans.

We discovered during the study that ``dose reduction'' permeated practically every step of CT-IQA's development. The ultimate goal of researchers is to decrease radiation exposure to patients while maintaining image quality, preserve human health to the greatest extent possible, and save medical costs.

On the one hand, researchers have created detectors with improved detection efficiencies and reconstruction algorithms that are more closely aligned with hardware characteristics in order to generate low-dose and high-quality CT images. To balance the image quality and the patient's radiation danger, medical professionals and technicians need to keep refining and optimizing the scanning parameters based on the patient's weight and the scan site. On the other hand, researchers are continuously investigating ways to maximize the utilization of low-dose CT images into high-quality images through the application of image post-processing algorithms. The dosage-performance of MBIR and ASIR was compared by Samei~{et al.}~\cite{71}, and the findings indicated that by using projected data, MBIR could lower the CT dose by two times. For certain jobs, there can be varying degrees of improvement in the quality of CT scans while the dose is maintained constant.

\subsubsection{Virtual Monoenergetic Reconstruction in CT-IQA}

With the constant advancement of CT equipment technology, contemporary CT scanners are now able to provide dual-energy imaging, that is, simultaneous scanning employing two distinct energy levels. Virtual single energy reconstruction is based on this type of dual energy imaging. It simulates imaging at a single energy level by processing raw data from CT scanners using specialized algorithms. This enhances contrast noise ratio, lessens beam hardening and metal artifacts, enhances the ability to detect anomalous structures, and improves overall image quality, giving physicians access to more useful data.

The subject of Photon Counting Detector (PCD) CT is the focus of the top three most referenced papers in this cluster. A novel energy-resolved X-ray detector is used by PCD-CT to provide high-quality CT images with greater resolution and contrast while avoiding electronic noise~\cite{72}. Consequently, PCD-CT can minimize radiation exposure, maximize contrast agent usage, and open doors for quantitative imaging~\cite{73}. Leng~{et al.}~\cite{74} summarized the principle, current situation, and application of PCD-CT. A thorough overview of the field is given via the inclusion of phantom, animal, and patient study examples. It is anticipated that the benefits of PCD-CT's high resolution, low dose, multi-energy, soft tissue, and low noise levels will significantly impact clinical diagnosis and raise diagnostic precision and clinical imaging standards.

\subsection{Prospective}

Finally, in order to make a scientific prediction about the direction of future development, we applied Structural Variation Analysis (SVA) to the themes following clustering (the principle is shown in~\cite{30}). We examined papers with citations spanning five years, as seen in Figure~\ref{Fig5}.

In cluster \#2, we observed that Viry~{et al.}~\cite{75} created a novel relationship between CT-IQA, reconstruction techniques, and dose by citing papers~\cite{76} and~\cite{77} to investigate the potential of various reconstruction algorithms in dose reduction. Nevertheless, links by themselves do not cause a meaningful paradigm change, so we do not conduct further analysis.

In cluster 1, there are three sub-clusters (sub-clusters 1, 2, and 3, from left to right). Paper~\cite{78} is considered the central node in sub-cluster 1, and papers~\cite{45} and~\cite{69} are cited in papers ~\cite{60, 79, 81} at the same time. The impact of various parameters and deep learning reconstruction algorithms on CT image quality is investigated, and a link between various reconstruction algorithms and CT system assessment guidelines is established. The findings demonstrate the significance of the reconstruction algorithm in the assessment of CT systems.

Within sub-cluster 2, researchers carried out research from various perspectives pertaining to the two concepts of ``deep learning'' and ``reconstruction algorithm'', as well as five highly cited papers ~\cite{67, 68, 70, 82, 86}. The performance comparison of several deep learning reconstruction techniques with conventional algorithms was examined in papers ~\cite{89, 90, 91, 92}. Park~{et al.}~\cite{93}examine the variation in image quality between the reconstructed standard-dose CT and the low-dose CT denoised by deep learning method. 

Four highly cited papers~\cite{95, 96, 97, 98} in sub-cluster 3 examined the performance of DLIR and contrasted it with two significant conventional reconstruction techniques, FBP and ASIR. Papers ~\cite{99, 100, 101} cite these works as a basis for their comparative analysis of deep learning reconstruction algorithms and conventional reconstruction techniques. Four conventional techniques are compared to a deep DLIR algorithm created especially for CCTA by Nagayama~{et al.}~\cite{100}. The outcomes demonstrate that the DLIR method may lessen artifacts and reconstruction time while also enhancing image quality. 

Sub-clusters 2 and 3 delve into the central theme of DLIR, while another cluster pertaining to the issue of ``Deep learning-based reconstruction algorithms'' has been formed from a variety of connections. Nevertheless, our analysis shows that there hasn't been a significant paradigm shift in DLIR research, and more investigation is still required. Current DLIR research focuses on the creation of novel algorithms and their performance comparison with conventional algorithms. 

\begin{figure*}[!t]
\centerline{\includegraphics[width=1\linewidth]{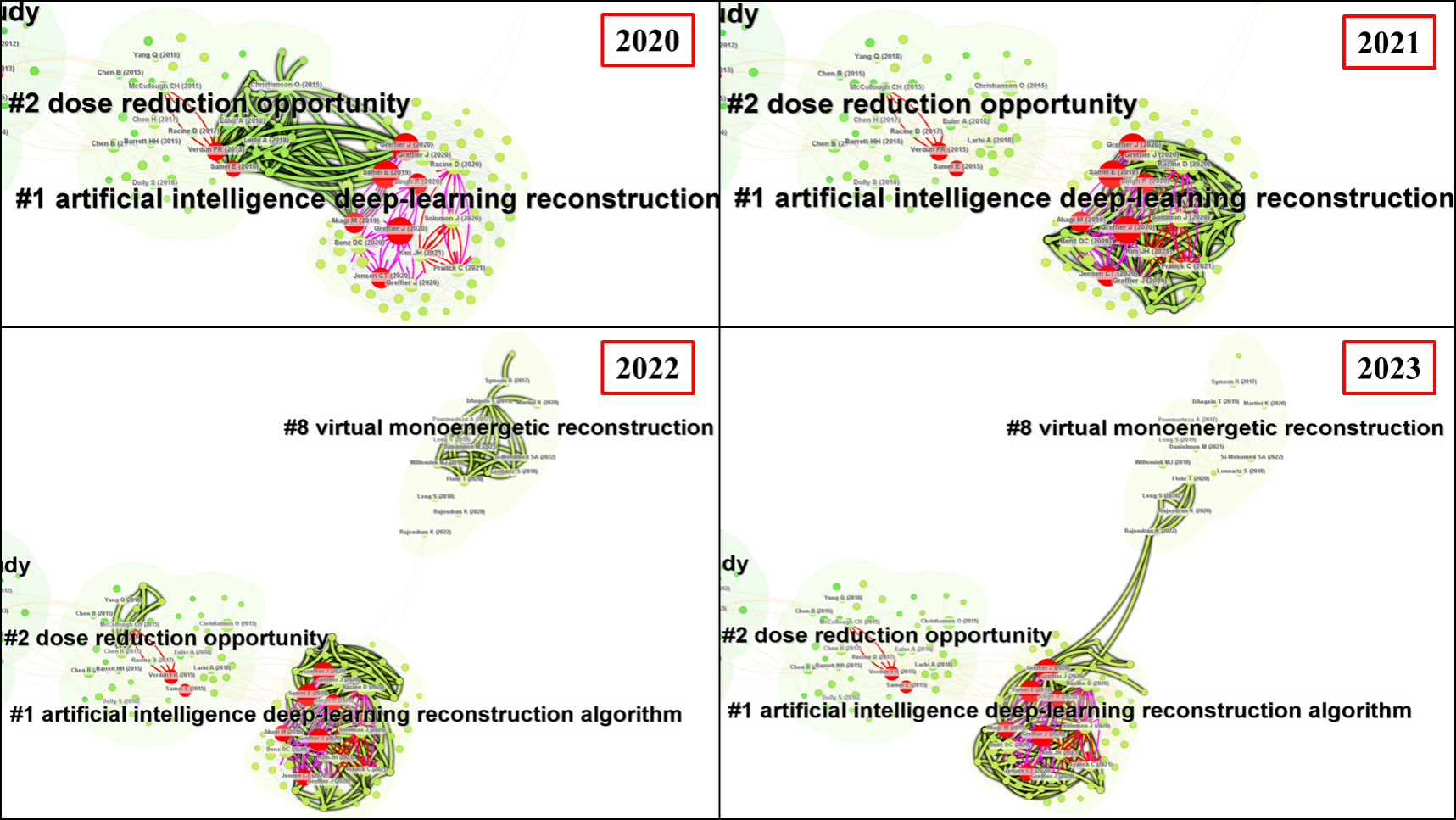}}
\caption{Link-walkthrough of the co-cited graph from 2020 to 2023 (the green connection line in the figure is the link-walkthrough path).}
\label{Fig7}
\end{figure*}

Furthermore, we examined the trajectory and evolution of the co-citation connection in link walkthrough the backdrop of the literature co-citation graph. Through path observation, we are able to comprehend the gradual evolution of CT-IQA and the changing interests of researchers, allowing us to investigate important advancements in the field as well as new research trends and associations. A link-walkthrough schematic from recent years is shown in Figure~\ref{Fig7}. 

Reconstruction techniques based on AI and deep learning started to offer new prospects for dose reduction in 2020, as shown in Figure~\ref{Fig3}, and a large number of related studies started to be employed in CT-IQA. Cluster \#2 and cluster \#1 also started to yield a big number of associations. The researchers have since turned their attention to cluster \#1. As we discussed in section 3.2.4, some researchers will begin focusing on virtual monoenergetic reconstruction in 2022. PCD-CT and other technologies are gradually making their way into the research field, and in 2023 they will start to connect with DLIR. As a result, academics will increasingly focus on creating novel CT imaging methods, reconstruction algorithms, and efficient CT-IQA. 

At the same time, we discovered that ``CT-IQA based on AI'' is starting to have a sustained development momentum and is anticipated to become a future research hotspot. This notion is also strongly supported by the present surge of key phrases and the cited paper in Figure~\ref{Fig4} and Figure~\ref{Fig6}. Simultaneously, research indicates that the AI-based CT-IQA model is anticipated to outperform numerous conventional methods. We anticipate that further research on AI-based CT-IQA will surface in the future and are optimistic that this area of study will develop into a new research hotspot.

\section{Discussion}

In the 1970s, CT imaging technology began to be used in clinical diagnosis. This has prompted the society to pay attention to the quality of CT images. CT-IQA could assist relevant personnel in doing quality assurance and enhancement. Researchers have raised standards for equipment performance and image processing algorithms as a result of the advancement of digital imaging technology, and CT-IQA is encountering both new opportunities and challenges.

Medical imaging experts or physicians subjectively assess medical images in the early stages of CT-IQA development using their professional knowledge and personal experience (clarity, contrast, sharpness, etc.). The results of the subjective assessment are uneven, the process is time-consuming and arduous, and it is easily influenced by personal subjective factors (such as working hours, mood, or clinical experience). The medical community is start to create norms and standards for image quality in order to help physicians make more accurate subjective assessments in order to address this issue. Simultaneously, an objective assessment of CT image quality started to take shape. The algorithm automatically finished the objective assessment, producing reliable and repeatable results and converting the image quality into quantifiable numbers for simple comparison and analysis. However, algorithms for objective assessment usually need to deal with large amounts of data and complex calculations. At the same time, due to the complexity of the human visual system, the existing algorithms still cannot fully simulate the subjective assessment.

The use of AI technology in CT-IQA has grown in the last few years. For instance, assessing the reconstruction algorithm, identifying artifacts, judging body position, and so forth. The automatic assessment of CT image quality can be achieved by training the AI model, increasing the assessment's efficiency and objectivity. Simultaneously, AI learns to continuously optimize and enhance its assessment model, bringing the assessment results closer to reality. In light of the research and analysis we have already done, the development of AI-based CT-IQA techniques is progressing quickly and is now a necessary trend in the field.

\subsection{Challenge}

At present, there are still many challenges in the CT-IQA model based on AI.

\subsubsection{Data}

A vast amount of high-quality annotated data is needed for AI learning, yet there isn't presently a sizable public CT-IQA data set available. There aren't many publicly accessible CT-IQA data sets at the moment. By introducing noise to CT scans, the majority of the current data sets are created and simulated. Although the quality of the simulation data is easily distinguishable, it is not able to completely model the complicated lesions and pathological circumstances that exist in the clinic, nor can it fully reflect the physiological distinctions between genuine patients.

On the one hand, manual data cleansing, screening, standardization, and labeling for private data sets is expensive and time-consuming. On the other hand, the collecting of data sets for specialized medical images necessitates rigorous ethical review and privacy protection. Data security and privacy protection concerns must also be taken into account. 

\subsubsection{Performance}

Complex model training and optimization procedures are required for AI models. Most IQA existing models are unable to manage varied and multi-center data due to the complexity and diversity of clinical contexts. They also have weak robustness, low generalization ability, and trouble expanding in practical applications. It is now challenging to outperform clinicians in the decision-making process. 

\subsubsection{Ethics and law}

In the context of healthcare, AI models struggle to explain their IQA decision-making processes, which makes the models challenging to apply in real-world scenarios and could present certain ethical and legal challenges. It becomes difficult to assign blame and establish legal liability when AI systems result in mistakes, mishaps, or injuries. As of right now, the legal system is still progressively adjusting to this new technology by creating relevant laws and frameworks for legal liability~\cite{102, 103}. Global interdisciplinary collaboration will be necessary to address these issues, involving the combined efforts of engineers, legal professionals, ethicists, and legislators to guarantee that the creation and use of AI conforms with moral principles, legal requirements, and regulatory frameworks.

\subsection{Opportunity and direction}

\subsubsection{AI-based CT-IQA Models}

Large model~\cite{106}, cloud computing~\cite{105}, and big data~\cite{104} technologies have emerged, and they can efficiently process and store vast amounts of medical image data in addition to offering strong processing power to support the development and use of intricate deep learning models. In order to address the ``black box'' issue with AI, researchers are also working hard to create new tools and techniques~\cite{a, b} that will make it easier for people to comprehend and explain how AI models make decisions and increase the explainability and transparency of AI models~\cite{107}. Simultaneously, moral ethics and associated legal difficulties~\cite{c} have drawn more attention~\cite{108}.  We have good reason to believe that as technology advances, AI-based CT-IQA models will become more and more significant in the CT-IQA domain and have a significant influence on the advancement of medical imaging.

\subsubsection{Scanners and Post-processing Algorithms}

Furthermore, we discovered during the course of our investigation that the dose of radiation associated with CT scanning varies based on various clinical downstream duties. The quality of CT images is mostly determined by the radiation dose, whether it is high or low. For example, large-scale population screening with low-dose CT can effectively increase cancer survival~\cite{109}, however the majority of the images are of low quality. Although there is a certain radiation risk associated with high-dose CT, it is appropriate for many clinical applications such as routine diagnosis, surgical planning, interventional therapy, etc. Enhancing the precision of CT scanning equipment (scanning speed, resolution, and dose control technology), enhancing the impact of the CT system reconstruction algorithm, and creating and refining CT image post-processing algorithms will also become crucial research areas because it is challenging to raise the quality of low-dose CT images on a large scale.

\subsubsection{Standardized CT-IQA System}

In order to establish a standardized CT-IQA system, researchers should also be dedicated to using subjective consensus, objective, and AI assessment methods. This includes creating unified assessment standards and metrics as well as reasonable assessment processes and methods. This will allow for a more thorough and objective assessment of the quality of CT images.

\section{Conclusion}

The paper reviews the development history, research focus, and potential future trends of CT-IQA. Research subjects that are popular right now include ``phantom study'', ``artificial intelligence deep-learning reconstruction algorithm'', ``dose reduction opportunity'', and ``virtual monoenergetic reconstruction''. Although AI-based CT-IQA has a bright future, there are obstacles to overcome, like limited explainability and limited data availability. Researchers have to carefully assess and strike a balance between data security, technology, ethics, and regulations. Simultaneously, boosting the effectiveness of the CT system reconstruction algorithm, creating a trustworthy CT image post-processing algorithm, and increasing the resolution and scanning speed of CT scanning apparatus will all become crucial research areas. Furthermore, academics ought to strive toward creating a CT-IQA framework that makes use of subjective consensus, objective, and AI assessment methods, and ongoing efforts to standardize CT-IQA.

We believe that AI-based medical IQA has excellent potential for use in clinical settings: 1) Managing the image quality manually has become a hard process due to the rise in popularity of medical examinations and the resulting growth in the quantity of images. Physician workloads can be significantly reduced and assessment efficiency can be increased with AI-assisted automated image assessment. 2) AI has the ability to standardize quality assessment criteria amongst various medical facilities and encourage them to identify one another's images, which lowers the amount of unnecessary tests and associated costs. 3) AI quality assessment can help medical facilities enhance the caliber of their medical imaging by offering more precise and comprehensive suggestions based on several factors like contrast, sharpness, artifacts, etc.

\appendices

\section{\textit{CiteSpace}}

\textit{CiteSpace} is an academic paper visualization analysis tool grounded in the principles of knowledge graph and scientometrics. It utilizes detailed information from collected paper as nodes (including keywords, references, authors, institutions, and countries), and co-occurrence relationships as connections to generate knowledge graphs. These graphs visually represent the structure, patterns, and distribution of knowledge within a field. Through interactive manipulation of the knowledge graphs produced by \textit{CiteSpace}, such as identifying nodes with high frequency and high betweenness centrality, searching for nodes with significant bursts, clustering co-occurring nodes, adding timelines to co-occurrence graphs, and incorporating structural variation analysis, an in-depth visual analysis is facilitated. \textit{CiteSpace} is applicable to a wide range of academic fields, aiding researchers in uncovering and exploring the knowledge structure, development trends, and research hotpots of their fields of interest. It also provides valuable insights into predicting future trends in these fields. The key terms based on which \textit{CiteSpace} software technology is used for visual analysis and their descriptions are shown in Table~\ref{tab4} below.

\begin{table*}
\caption{\textit{CiteSpace} software key terms and their description.}
\label{table}
\setlength{\tabcolsep}{3pt}
\begin{tabular}{|p{50pt}|p{450pt}|}
\hline
\textbf{Metric} & \textbf{Description} \\ \hline
Node & In a knowledge graph, nodes are the basic units that represent entities or concepts. In \textit{CiteSpace}, nodes can be authors, institutions, countries, keywords, terms, categories, references, cited authors, cited journals, and so on. Nodes are connected by edges (relationships) to form the structure of the graph. \\ \hline
Keyword co-occurrence & Keyword co-occurrence refers to two or more keywords that appear simultaneously in the same document. In the knowledge graph, this co-occurrence relationship can be used to reveal possible correlations between these keywords. \\ \hline
Co-Citation & When two or more documents are cited by the same document, they are considered to be co-cited. In the knowledge graph, the co-citation relationship can be used to reveal the similarity, correlation or common research area between literatures. \\ \hline
Frequency & In the knowledge graph, frequency usually refers to the number of times a node occurs in the whole graph. Nodes with high frequency usually indicate importance or significance. \\ \hline
Betweenness Centrality & Betweenness centrality is a measure of how well a node is connected to other nodes in a network. Nodes with high betweenness centrality play an important role as bridges in information propagation because they are located on short paths in the network. \\ \hline
Burst Detection & Burst detection is used to identify events or relationships that occur with an unusually high frequency within a designated time frame. In the knowledge graph, it helps to discover concepts that cause concern or change in a specific time period. \\ \hline
Clustering & Clustering refers to the method of grouping nodes within a graph based on certain similarities or connectivities. This process aids in deciphering substructures and groups within the knowledge graph, thereby underlying patterns and associations. \\ \hline
Cluster Label & A cluster label is a descriptive label for a cluster or group that generalizes the topic or feature of the nodes in the cluster. \\ \hline
Structural Variation Analysis (SVA) & SVA involves examining alterations in the structure of a graph, encompassing the identification of changes in nodes and edges, and comprehending the implications of these alterations on the graph's entirety. This approach is instrumental in apprehending the dynamic characteristics and evolving trends of the knowledge graph.\\ \hline

\hline

\end{tabular}
\label{tab4}
\end{table*}

\section*{Acknowledgment}

This work is supported by Science and Technology Development Fund of Macao (0021/2022/AGJ), Macao Polytechnic University Grant (RP/FCA-15/2022), and Macao Polytechnic University Grant (RP/FCSD-01/2022).

\section*{References}
\bibliographystyle{IEEEtran}
\bibliography{refs}

\end{document}